\documentclass[a4paper,10pt,twoside]{cpc-hepnp}

\usepackage{multicol}
\usepackage{graphicx}
\usepackage{booktabs}
\usepackage{amssymb,bm,mathrsfs,bbm,amscd}
\usepackage[tbtags]{amsmath}
\usepackage{lastpage}

\begin{document}

\fancyhead[co]{\footnotesize D. Werthm\"uller: Investigation of the anomaly in $\eta$-photoproduction off the neutron}

\footnotetext[0]{Received 7 August 2009}

\title{Investigation of the anomaly in \\ $\eta$-photoproduction off the neutron\thanks{Supported by Schweizerischer Nationalfonds, DFG, and EU/FP6}}

\author{%
      D. Werthm\"uller$^{1;1)}$\email{dominik.werthmueller@unibas.ch} (for the Crystal Ball/TAPS collaborations)
}
\maketitle

\address{%
1~(Department of Physics, University of Basel, Klingelbergstrasse 82, CH-4056 Basel, Switzerland)\\
}

\begin{abstract}
Quasi-free photoproduction of $\eta$-mesons off the neutron and off the proton has been studied using a deuterium target and bremsstrahlung photons produced by MAMI-C with incident energies up to 1.5 GeV. The $\eta$-mesons were detected in coincidence with the recoil nucleons thus a fully exclusive measurement was performed. Preliminary results show a bump-like structure in the excitation function for the neutron close to \mbox{W $\approx$ 1675 MeV} which is not seen for the proton. Considering the experimental resolution and using a Breit-Wigner fit the width of this structure was approximated below 50 MeV.
\end{abstract}



\begin{multicols}{2}

\section{Introduction}
Photoproduction of mesons is an excellent tool to investigate nucleon resonances. Studying reactions on both the proton and the neutron can help to reveal the isospin structure of resonances. In case of $\eta$-photoproduction only N$^*$ resonances can be excited because of the isospin zero of the $\eta$-meson and can therefore be studied separately from the $\Delta$ resonances. On the proton $\eta$-photoproduction was studied in great detail and a strong dominance of the S$_{11}$(1535) in the threshold region was found\cite{eta_p_1, eta_p_2, eta_p_3, eta_p_4, eta_p_5, eta_p_6, eta_p_7, eta_p_8, eta_p_9, eta_p_10, eta_p_11, eta_p_12}. In case of the neutron one has to use nuclear targets as deuterium or helium-3/4 because a free neutron target is not available. In the analysis it is then necessary to account for possible nuclear effects (e.g. FSI, Fermi motion) which could have an unwanted influence in the result. With these measurements\cite{eta_n_1, eta_n_2, eta_n_3, eta_n_4, eta_n_5, eta_n_6, eta_n_7, eta_n_8} the isospin structure of the electromagnetic excitation of the S$_{11}$(1535) was found to be mostly iso-vector leading to a ratio of 2/3 for the cross sections of the neutron and the proton. Beyond the S$_{11}$(1535), models suggest a larger ratio due to higher lying resonances that couple more strongly to the neutron than to the proton. A chiral soliton based model\cite{chiral_soliton_pentaquark} even predicts the existence of a narrow P$_{11}$(1680) which is the non-strange member of the anti-decuplet of pentaquarks. Experimentally, a bump-like structure in the excitation function of the neutron in the suggested energy region was discovered by the GRAAL collaboration\cite{prev_res_graal} and later confirmed by LNS\cite{prev_res_lns} and the CBELSA/TAPS collaboration\cite{prev_res_cbelsa}. In case of the proton no equivalent structure was seen (hence the name ``neutron anomaly'').

Here we report from very preliminary results of a new experiment using a deuterium target. This measurement at the Mainz MAMI accelerator provides higher statistics, especially for $\eta$-mesons at backward angles, which were suppressed in the CBELSA/TAPS experiment due to trigger conditions. The aim is to establish an upper limit for the width of the structure and to improve the quality of the differential cross section data.

\begin{center}
\includegraphics[width=\columnwidth]{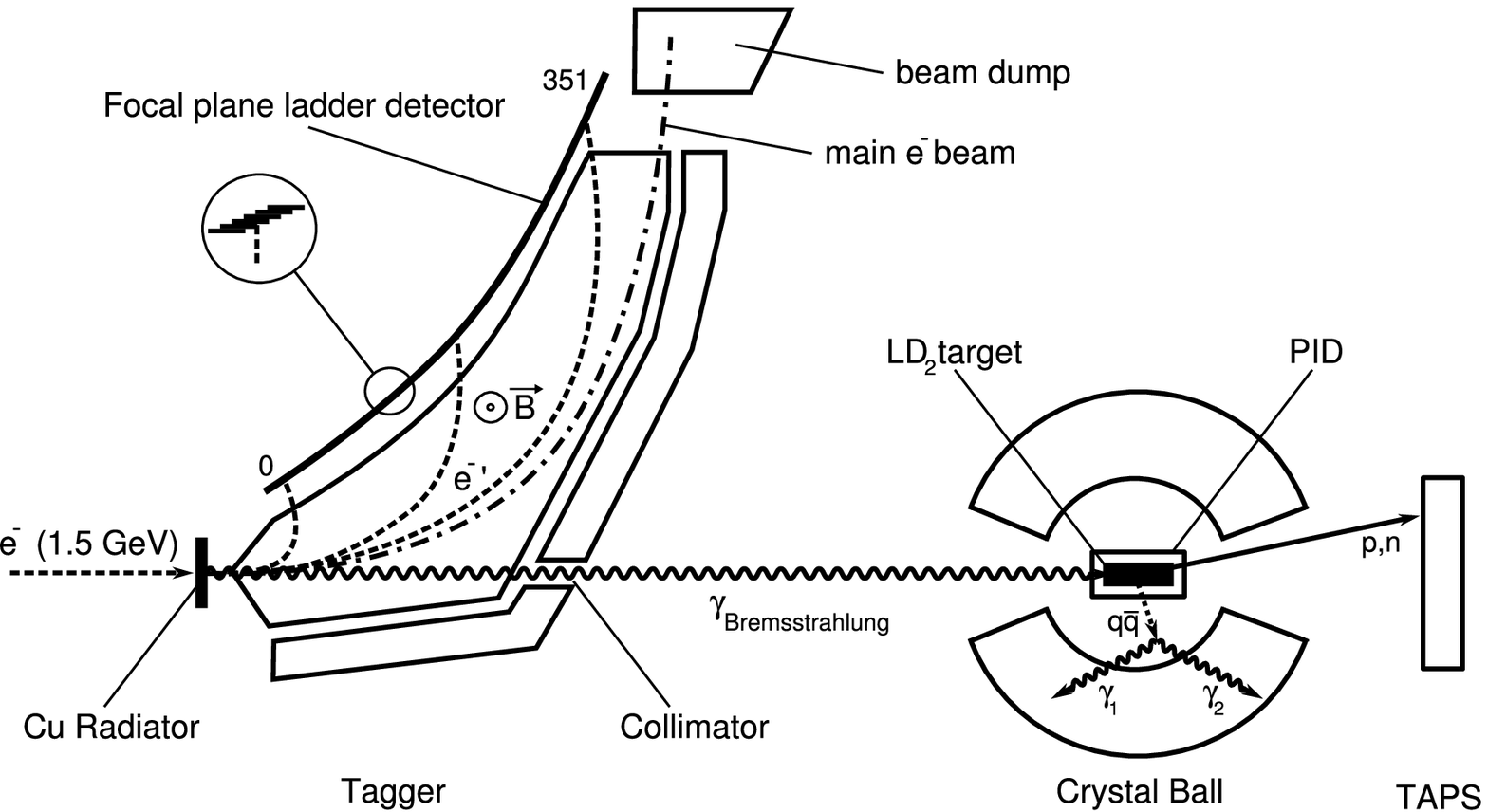}
\figcaption{\label{fig:exp_setup} Overview of the experimental setup}
\end{center}

\section{Experiment}
The measurement was performed at the electron accelerator facility MAMI\cite{MAMI,MAMI_C} in Mainz. A photon beam was produced from a 10 nA electron beam of 1.5 GeV energy via bremsstrahlung using a 10 $\mu$m copper radiator. The energy of the photons was determined with a momentum analysis of the scattered electron in a magnetic spectrometer (Glasgow photon tagger\cite{Tagger_1, Tagger_2, Tagger_3}). After collimation the beam impinged on the liquid deuterium target of 4.76 cm length. The target was surrounded by a cylindrical plastic scintillator strip detector\cite{PID}, which was used for charged particle identification, and the spherical electromagnetic calorimeter Crystal Ball\cite{CB}. This detector consists of 672 NaI crystals and is covering 94\% of 4$\pi$ steradians. The hole in forward direction of Crystal Ball is closed by the TAPS detector\cite{TAPS_1, TAPS_2} which is made of 384 BaF$_2$ crystals. In front of every crystal a thin plastic scintillator element is installed as a charged veto detector. As trigger condition a deposited energy sum of 300 MeV in the Crystal Ball and a total multiplicity of two or more hits in both calorimeters was requested.

\section{Analysis}
Events of the quasi-free reactions $\gamma p\rightarrow \eta p$ were selected by looking for exactly two neutral hits (i.e. hits with no veto signature) in the detectors, coming from the two photons of the $\eta\rightarrow 2\gamma$ decay channel, and exactly one additional charged hit of the proton. In case of the quasi-free reaction on the neutron $\gamma n\rightarrow \eta n$ exactly three neutral hits were requested in a first step. Afterwards the invariant mass for all two particle combinations was calculated assuming two photons. Via a $\chi^2$-minimization the best combination to form a $\eta$-meson was selected and the remaining particle was identified as the neutron. For both reactions an invariant mass cut requesting \mbox{480 MeV $< m_{\gamma\gamma} <$ 620 MeV} was applied to clearly select events with a $\eta$-meson in the final state. Background rejection was improved by the condition that $\eta$-meson and recoil nucleon were approximately coplanar. A cut was applied at \mbox{130$^{\circ}$ $< |\Delta\Phi| <$ 220$^{\circ}$}. Finally, assuming $\eta$-photoproduction off a free nucleon at rest, the missing mass can be calculated. Although the observed peak at the nucleon mass is broadened by Fermi motion a cut that depends on the energy of the incident photon can be applied on it which helps to remove background mainly coming from $\eta\pi$-photoproduction. 

Random coincidences between the photon tagging spectrometer and the other detectors are leading to background events. They were removed using the common method of statistical subtraction where events of a true random coincidence time interval are subtracted after normalization from events in the true coincidence time interval.

\begin{center}
\includegraphics[width=\columnwidth]{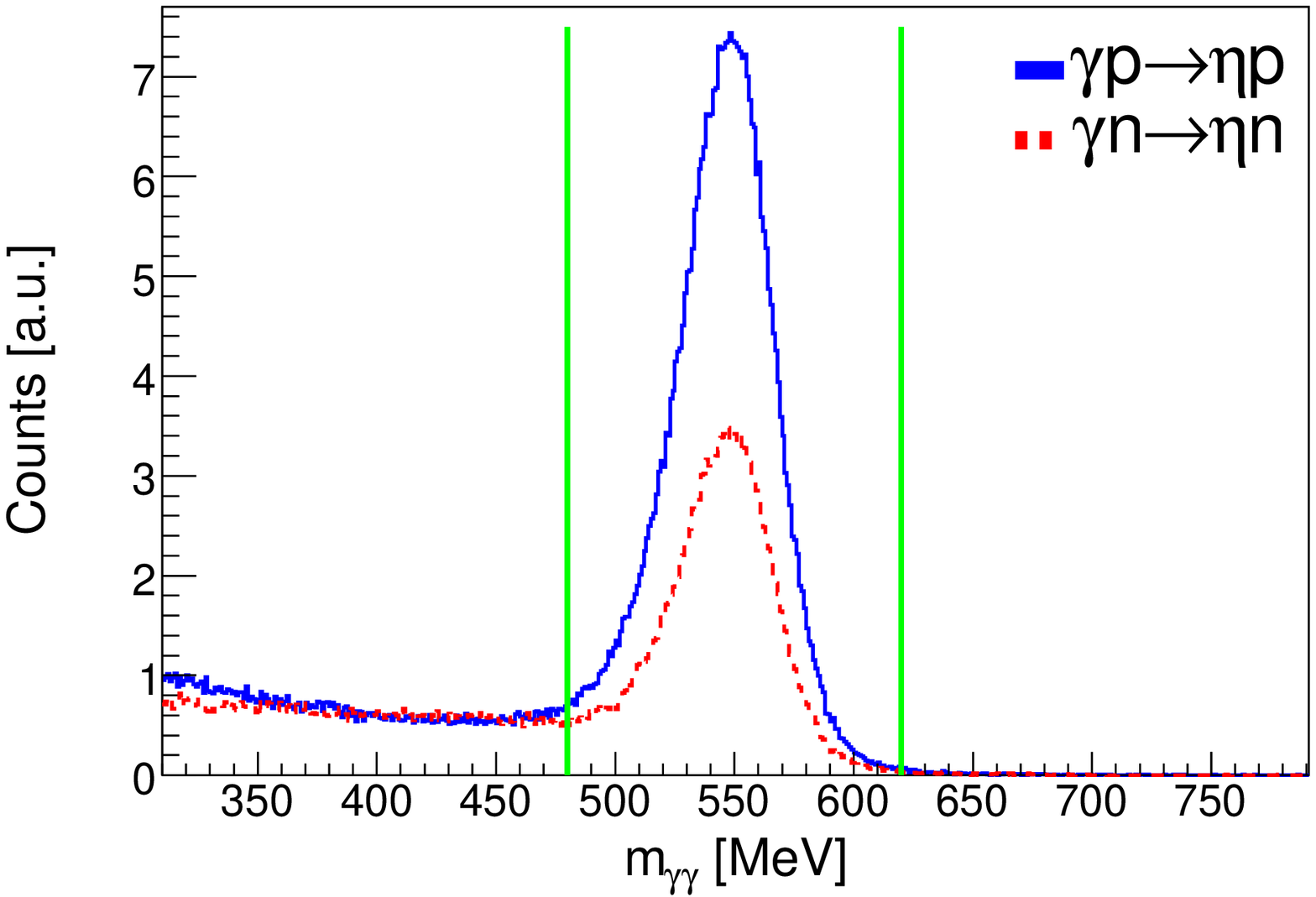}
\figcaption{\label{fig:inv_mass} Invariant mass histogram of two photons after the missing mass and the coplanarity cut: Solid blue line: for the quasi-free reaction on the proton. Dashed red line: for the quasi-free reaction on the neutron. The green vertical lines denote the invariant mass cut limits to select events with an $\eta$-meson.}
\end{center}

The excitation function values for a certain beam energy bin were derived by integrating the number of events detected per $\cos(\theta_{CM})$ bin over the full solid angle. The energy and angle dependent detection efficiencies that are normally estimated by Monte-Carlo simulations were not yet determined. The flux of the photon beam per energy bin was calculated as the product of the number of scattered electrons in the corresponding channel of the photon tagging spectrometer and the tagging efficiency of this particular channel. The tagging efficiency is the ratio of the number of photons reaching the target after beam collimation and the detected electrons in the tagging spectrometer. It was measured on a daily basis in separate measurements using a very low intensity beam and a special detector that was driven directly into the photon beam counting single photons with almost 100\% efficiency. 

The energy in the center-of-mass system $W$ was calculated in two different ways: First, assuming the reaction took place on a free nucleon at rest, $W_{beam}$ can be directly calculated with the energy of the incoming photon:
\begin{equation}
\label{eqn:W_beam}
W_{beam} = m(\gamma N) = \sqrt{2E_{\gamma}m_{N} + m_{N}^{2}}.
\end{equation}
This is not the true center-of-mass energy at which the reaction occurred as the nucleons are not at rest inside the deuteron nucleus but are moving with a certain Fermi momentum. Considering this it is necessary to calculate the true energy 
\begin{equation}
\label{eqn:W_true}
W_{true} = m(\eta N') = \sqrt{(E_{\eta} + E_{N'})^2 - (\vec{p}_{\eta} + \vec{p}_{N'})^2}. 
\end{equation}
using the 4-momenta of the final state $\eta$-meson and the recoil nucleon. Unlike for photons there is no direct relationship for neutrons between the deposited energy in the detector and the kinetic energy of the particle. While for protons a Monte-Carlo simulation based calibration up to the detector punch-through energy can be established, no such technique can be used for neutrons, which are depositing their energy via nuclear reactions. Therefore the kinetic energy of the recoil nucleons has to be determined in a different way. Knowing the incident photon energy, the particle masses, the 4-momentum of the $\eta$-meson and the direction of the recoil nucleon the kinematics is fully determined and the kinetic energy of the recoil nucleon can be calculated. The high granularity of the detectors ensures a good position and consequently a good energy resolution when using this method.

\end{multicols}
\ruleup
\begin{center}
\includegraphics[width=\textwidth]{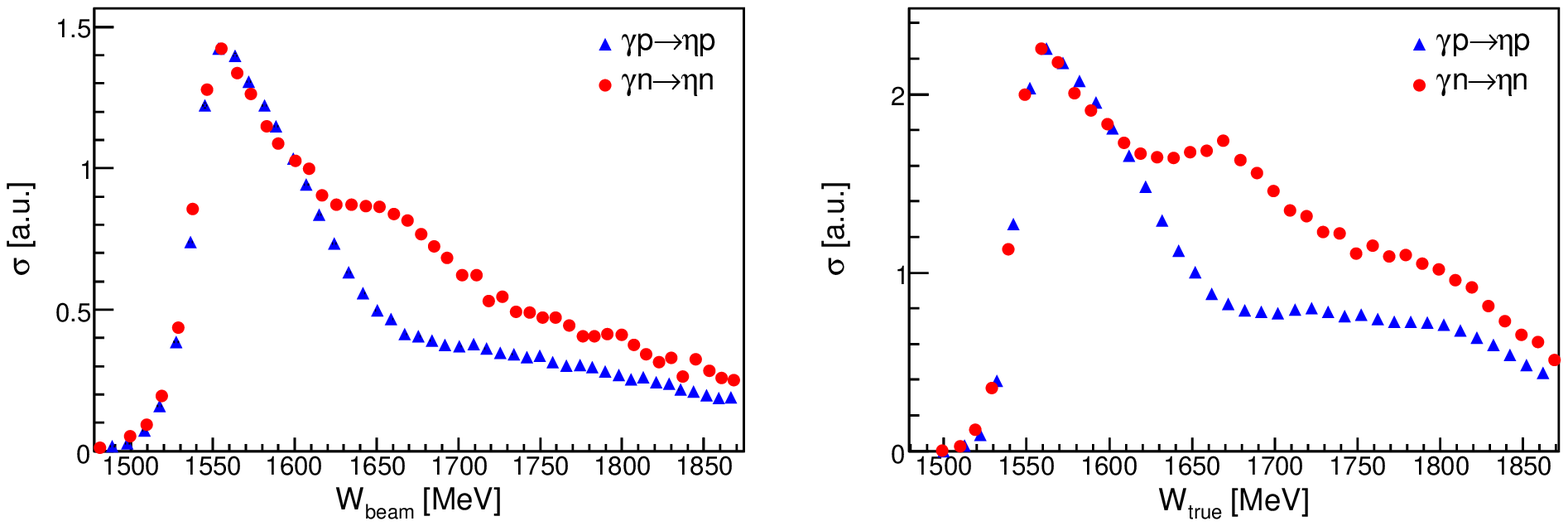}
\figcaption{\label{fig:exfunc_soft_cop} Very preliminary excitation functions using a 130$^{\circ}$ $< |\Delta\Phi| <$ 220$^{\circ}$ coplanarity cut: Left-hand side: center-of-mass energy calculated assuming a free nucleon. Right-hand side: center-of-mass energy calculated using the final state $\eta$ and recoil nucleon 4-momenta. Blue triangles: proton data. Red circles: neutron data. The excitation function of the proton is scaled to the one for the neutron in the maximum of the S$_{11}$(1535). All data have not been corrected for detection efficiency.}
\includegraphics[width=\textwidth]{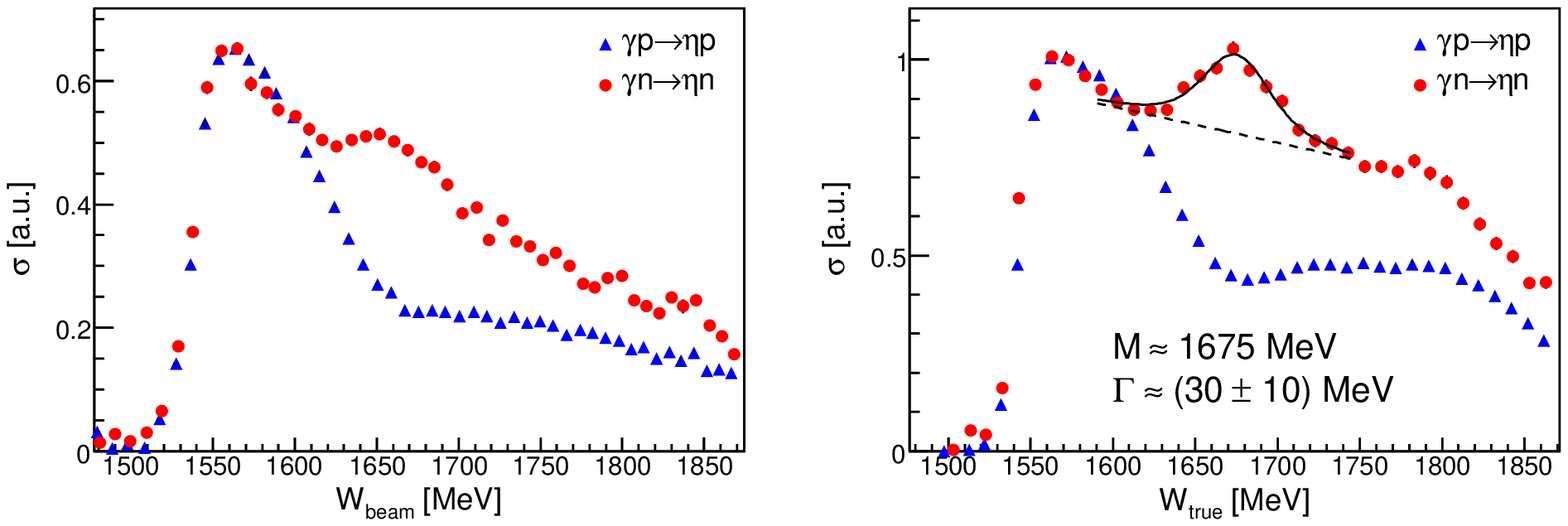}
\figcaption{\label{fig:exfunc_hard_cop} Very preliminary excitation functions using a 170$^{\circ}$ $< |\Delta\Phi| <$ 190$^{\circ}$ coplanarity cut: Notation and symbols are the same as in Fig. \ref{fig:exfunc_soft_cop}. Curves on the right-hand side: Solid: first-order polynomial + Breit-Wigner-function convoluted with a Gaussian. Dashed: first-order polynomial background function.}
\end{center}

\ruledown
\begin{multicols}{2}

\section{Very preliminary results}
Fig. \ref{fig:exfunc_soft_cop} shows the excitation functions using a soft coplanarity cut. The center-of-mass energy was calculated using the two methods described in the previous section: On the left-hand side the excitation function is shown in dependance of $W_{beam}$ while on the right-hand side it is depending on $W_{true}$. In both cases a clear difference between the proton and the neutron case starting around 1600 MeV is visible. If the true center-of-mass energy is calculated (right-hand side) a bump-like structure is emerging and peaks close to 1670 MeV. To improve the resolution in the center-of-mass energy a more strict coplanarity cut requesting \mbox{170$^{\circ}$ $< |\Delta\Phi| <$ 190$^{\circ}$} can be applied that removes events where the participating nucleon had a large Fermi momentum. The corresponding excitation functions are shown in Figure \ref{fig:exfunc_hard_cop}. Even in the excitation function depending on $W_{beam}$ a bump is now clearly seen whereas a distinct peak appears in the $W_{true}$-dependent excitation function. 

To estimate the true width of this structure the data was fitted using the sum of a first-order polynomial and a Breit-Wigner-function that was convoluted with a Gaussian to account for the experimental resolution. An experimental resolution of 30 MeV was determined by simulating the decay of a resonant state with fixed energy at 1680 MeV into a $\eta$-meson and a neutron within a Geant4\cite{geant4_1, geant4_2} based model of the detector setup. As a very preliminary result the following values were obtained for the Breit-Wigner center $M$ and width $\Gamma$ of the structure: 
\begin{eqnarray}
\label{eqn:result_center}
M & \approx & 1675 \,\, \mathrm{MeV} \\
\Gamma & \approx & (30 \pm 10) \,\, \mathrm{MeV}
\end{eqnarray}

\section{Conclusion}
The preliminary results confirm the previous findings of the other experiments concerning the ``neutron anomaly'' in $\eta$-photoproduction. The significance will be increased by the final results that will also include angular distributions. The nature of this structure will be further investigated via the measurement of polarization observables, which are most sensitive to the quantum numbers of a possible excited nucleon resonance.

\end{multicols}

\vspace{-2mm}
\centerline{\rule{80mm}{0.1pt}}
\vspace{2mm}

\begin{multicols}{2}

\end{multicols}

\vspace{5mm}

\clearpage

\end{document}